\documentclass{aa}

\usepackage{graphicx}
\usepackage{txfonts}
\usepackage{lscape}
\usepackage{color}

\bibpunct{(}{)}{;}{a}{}{,}

\usepackage{natbib}
\usepackage{amstext}

\begin{document}


\title{{\it N}-body simulations of $\gamma$ gravity}

\author{Marcelo Vargas dos Santos\inst{1,2}\fnmsep\thanks{vargas@if.ufrj.br}
          \and
          Hans A. Winther\inst{3}\fnmsep\thanks{hans.a.winther@gmail.com}
          \and
          David F. Mota\inst{2}\fnmsep\thanks{d.f.mota@astro.uio.no}
          \and
          Ioav Waga\inst{1}\fnmsep\thanks{ioav@.if.ufrj.br}
          }

   \institute{Instituto de F\'{i}sica, Universidade Federal do Rio de Janeiro C. P. 68528, CEP 21941-972, Rio de Janeiro, RJ, Brazil\\
         \and
            Institute of Theoretical Astrophysics, University of Oslo, Postboks 1029, 0315 Oslo, Norway\\
         \and 
                Astrophysics, University of Oxford, DWB, Keble Road, Oxford, OX1 3RH, UK
             }

\titlerunning{{\it N}-body Simulations of $\gamma$ Gravity}
\authorrunning{M. Vargas dos Santos et al.}

\abstract{We have investigated structure formation in the $\gamma$ gravity $f(R)$ model with {\it N}-body simulations. The $\gamma$ gravity model is a proposal which, unlike other viable $f(R)$ models, not only changes the gravitational dynamics, but can in principle also have signatures at the background level that
are different from those obtained in $\Lambda$CDM (Cosmological constant, Cold Dark Matter). The aim of this paper is to study the nonlinear regime of the model in the case where, at late times, the background differs from $\Lambda$CDM. We quantify the signatures produced on the power spectrum, the halo mass function, and the density and velocity profiles. To appreciate the features of the model, we have compared it to $\Lambda$CDM and the Hu-Sawicki $f(R)$ models. For the considered set of parameters we find that the screening mechanism is ineffective, which gives rise to deviations in the halo mass function that disagree with observations. This does not rule out the model per se, but requires choices of parameters such that $|f_{R0}|$ is much smaller, which would imply that its cosmic expansion history cannot be distinguished from $\Lambda$CDM at the background level.}


\maketitle

\section{Introduction}

Since the discovery in 1998 \citep{Riess:1998cb,Perlmutter:1998np} that the Universe is speeding up instead of slowing down (as would be expected if gravity is always attractive), considerable effort has been devoted to understanding the physical mechanism behind this cosmic acceleration. The two main theoretical approaches considered in the literature to explain this phenomena are (1) to assume the existence of a new component with a sufficiently negative pressure ($p < -\rho/3$),  generically denoted dark energy, and (2) to consider that general relativity has to be modified at large scales, or, more accurately, at low curvature (modified gravity).  The simplest dark energy candidate is Einstein's cosmological constant ($\Lambda$) with an equation of state $w_{\rm DE}\equiv p_{\rm DE}/\rho_{\rm DE}=-1$. However, in spite of its very good accordance with current observations, $\Lambda$ has some theoretical difficulties such as its tiny value as compared with theoretical predictions of the vacuum energy density, the cosmic coincidence problem, and related fine-tuning. This situation has motivated the search for alternatives like modified-gravity theories. The simplest modified-gravity candidates are the so-called $f(R)$-theories, in which the Lagrangian density $\mathcal{L}=R+f(R)$ is a nonlinear function of the Ricci scalar $R$.

As is well known, metric $f(R)$-theories can be thought of as a special case of a scalar-tensor theory; a Brans-Dicke model with a coupling constant $\omega_{\rm BD}=0$. An accelerated expansion appears naturally in these theories. The very first inflationary model, proposed by Starobinsky more than three decades ago \citep{Starobinsky:1980te}, is driven by a term of the type $f(R)=\alpha R^2$ ($\alpha>0$) and is still in excellent accordance with observations \citep{Planck:2013jfk}.  More recently, the idea of an acceleration driven by late-time curvature has also been explored in \cite{Capozziello:2003gx} and \cite{Carroll:2003wy}. These authors considered a theory in which $f(R)=-\alpha R^{-n}$ ($n>0$ and $\alpha>0$). However, these models do not have a regular matter-dominated era and are incompatible with structure formation \citep{Planck:2013jfk}.

To build a cosmologically viable $f(R)$ theory, some stability conditions have to be satisfied \citep{Pogosian:2007sw}: (a) $f_{RR}\equiv d^2f/dR^2>0$ (no tachyons);  (b) $1 + f_{R}\equiv 1 + df/dR > 0$  [the effective gravitational constant, ($G_{\rm eff}=G_N/(1+f_R)$) does not change sign (no ghosts)]; (c) after inflation, $ \lim_{R\rightarrow\infty}f(R)/R=0$ and $ \lim_{R\rightarrow\infty}f_R=0$ (General Relativity is recovered at early times) ; (d) $|f_R|$ is small at recent times, to satisfy solar system and galactic scale constraints. In addition to these conditions, there are some desirable characteristics that a viable cosmological model has to satisfy \citep{Amendola:2006we}. It should have a radiation-dominated era at early times and a saddle-point matter-dominated phase followed by an accelerated expansion as a final attractor. By using the parameters  $\bar{m}\equiv Rf_{,RR}/(1+f_{R})\;$ and $\;\mathfrak{r} \equiv -R(1+f_{R})/(R+f)\,$, it can be shown that an early matter-dominated epoch of the Universe can be achieved if $ \bar{m}(\mathfrak{r} \approx -1) \approx 0^+$ and $\bar{m}/\mathfrak{r}(\mathfrak{r}\approx -1) > -1$. Furthermore, a necessary condition for a late-time accelerated attractor is $0<\bar{m}(\mathfrak{r}\approx -2)\leq 1$.

There are viable $f(R)$ gravity theories that satisfy all the criteria mentioned above \citep{Hu:2007nk,Starobinsky:2007hu,Appleby:2007vb,Cognola:2007zu,Linder:2009jz,O'Dwyer:2013mza}. However, there is a generic difficulty from which all these ``viable'' $f(R)$ theories \citep{Thongkool:2009js} suffer: the curvature singularity in cosmic evolution at a finite redshift \citep{Frolov:2008uf}. It can be shown that this type of singularity problem can be cured, for instance, by adding  a high-curvature term proportional to $R^2$ \citep{Appleby:2009uf} to the density Lagrangian. Therefore, it is not possible to have cosmic acceleration with a totally consistent $f(R)$ theory modifying gravity only at low curvatures. We remark that we do not address this problem here and only consider modifications at low curvatures.

We consider the specific case of a viable $f(R)$ theory called $\gamma$ gravity \citep{O'Dwyer:2013mza}. Generically, in almost all viable $f(R)$ theories, structure formation imposes such strong constraints on the parameters of the models that the effective equation of state parameter cannot be distinguished from that of a cosmological constant. In $\gamma$ gravity the steep dependence on the Ricci scalar $R$ facilitates the agreement with structure formation. \cite{O'Dwyer:2013mza} showed that, in principle, the parameter that controls the steepness in $\gamma$ gravity allows measurable deviations  from $\Lambda$CDM (Cosmological constant, Cold Dark Matter) at both linear perturbation and background levels, while still compatible with both current observations. The main goal of this paper is to study the effects of $\gamma$ gravity on the structure formation at nonlinear scales for choices of parameters where the model has observable signatures\footnote{With observable signatures we mean that the equation of state differs enough from the $\Lambda$CDM value of $w=-1$ at low redshifts that such a deviation could be detected by near-future experiments like WFIRST \cite{2015arXiv150303757S},
for example.} on the background expansion history of our Universe.

We go one step further and analyze the nonlinear evolution of structures computed from numerical simulations. The code that this paper is based on is a slight modification of {\tt ISIS} \citep{Llinares:2013jza}, which in turn is a modification of the {\tt RAMSES} hydrodynamic {\it N}-body code \citep{Teyssier:2001cp}. See also \cite{2011PhRvD..83d4007Z,2012JCAP...01..051L,Oyaizu:2008tb,li1,li2,li3,li4,2013MNRAS.436..348P} for other codes that have implemented and performed simulations of $f(R)$ gravity and \citep{2015arXiv150606384W} for a recent code comparison between these codes.

We only consider the modifications made to implement the $\gamma$ gravity field equations in the modified gravity part of the {\it N}-body code. For more details on the implementation of the scalar fields and other technicalities we refer to \cite{Llinares:2013jza}. For this purpose, we focus on simple observables such as the matter power spectrum, halo mass function, density profiles, and velocity profiles to investigate modified gravity signatures that were previously studied in \cite{Li:2011vk,Hammami:2015iwa,Schmidt:2008tn,Gronke:2014gaa,Lombriser:2011zw,Lombriser:2013eza,Winther:2011qb,Terukina:2012ji,Shi:2015aya,Pujol:2013yna,Li:2012by,Hellwing:2011ne,2013PhRvD..88j3507H,He:2015bua,2010PhRvD..81j3002S,2013JCAP...04..029B,2015JCAP...10..036T,2015MNRAS.449.2837G,2013MNRAS.428..743L}
that can also be observed \citep{2015arXiv150107274B,2015MNRAS.451.4215Z,2012PhRvD..85l3513M,2014arXiv1410.8355C,2015MNRAS.451.1036C,2015PhRvD..92d3522S,2015MNRAS.452.1171W,Jain:2010ka,Schmidt:2010jr,2014PhRvL.112v1102H}.


This paper is organized as follows: in Sect. \ref{model} we revisit the $\gamma$ gravity model and show the main properties of the background and linear perturbation evolution. Section \ref{nbody} details the dynamics equations of scalaron field ($f_R$) and particle movement equations for $\gamma$ gravity, which must be solved by our code during the simulations. The method for solving these equations is briefly explained in Sect. \ref{isis}, and the code is tested in Sect. \ref{test}. Finally our results are shown in Sect. \ref{results}, and we conclude in Sect. \ref{conc}.

\section{$\gamma$ gravity review} \label{model}

We investigate spatially flat cosmological models in the context of $\gamma$ gravity \citep{O'Dwyer:2013mza}, a viable $f(R)$ theory defined by the following ansatz:
\begin{align}
f(R) &= -\frac{\alpha R_*}{n}\gamma\left[\frac{1}{n}, \left(\frac{R}{R_*}\right)^n\right],
\label{fRgamma}
\end{align}
where $\gamma(n,x) = \int_0^x t^{n-1} e^{-t}dt$ is the incomplete $\Gamma$-function and $\alpha$, $n$ and $R_*$ are free positive constants. In reality, $\gamma$ gravity can be thought of as a simple generalization of exponential gravity \citep{Linder:2009jz}
\begin{equation}
f(R)=-\alpha R_{\ast} (1-e^{-R/R_{\ast}}),
\end{equation}
obtained by fixing $n=1$ in Eq.\,(\ref{fRgamma}). We emphasize that $\gamma$ gravity can satisfy all the stability and viability conditions. As discussed in \cite{O'Dwyer:2013mza}, for fixed $n$, there is a minimum value ($\alpha_{\rm min}$) of the parameter $\alpha$ such that for values $\alpha > \alpha_{\rm min}$ a late-time accelerated attractor is achieved. We consider this case throughout. From Eq.\,(\ref{fRgamma}) we obtain the following derivatives:
\begin{align}
&f_R = -\alpha e^{-\left(\frac{R}{R_*}\right)^n}\label{fReq},\\
&f_{RR} = \frac{\alpha n}{R} \left(\frac{R}{R_*}\right)^n e^{-\left(\frac{R}{R_*}\right)^n}.
\end{align}
We note from Eq.~({\ref{fReq}) that with increasing $n$, the steepness of the $f(R)$ function increases. Higher $n$ means smaller $|f_{R0}|,$ and the departures from GR will be smaller accordingly.

Although there is no cosmological constant, $f(0)=0$, it follows from Eq.~(\ref{fRgamma}) that GR with $\Lambda$ is recovered at high curvatures. Therefore, for $R \gg R_{\ast}$ the models behave like $\Lambda$CDM. Since we are mainly interested in phenomena that occurred after the beginning of the matter-dominated era, we neglect radiation and write the effective cosmological constant (the cosmological constant of the reference $\Lambda$CDM model) as
\begin{align}
\tilde{\Lambda} = \frac{\alpha R_*}{2n}\Gamma(1/n) = 3 \tilde{H}_0^2 (1-\tilde{\Omega}_{m0}).
\label{lambda}
\end{align}
In the equation above, $\tilde{\Omega}_{m0}$ denotes the present value of the matter density parameter that a $\Lambda$CDM model would have if it had the same matter density today ($\bar{\rho}_{m0}$) as the modified gravity $f(R)$ model. $\tilde{H}_0$ represents the Hubble constant in the reference $\Lambda$CDM model. Therefore, we have $m^2 \equiv 8\pi G\bar{\rho}_{m0}/3= \tilde{\Omega}_{m0}\tilde{H}_0^2=\Omega_{m0}H_0^2$, where $\Omega_{m0}$ and $H_0$ are the present value of the matter energy density parameter and Hubble parameter in the $f(R)$ model,
respectively. It is useful to rewrite $R_*$ as
\begin{eqnarray}
        \frac{R_*}{m^2} = \frac{6 \, n \, d}{\alpha \Gamma (1/n)},
        \label{R*}
\end{eqnarray}
where $d = (1-\tilde{\Omega}_{m0})/\tilde{\Omega}_{m0}$. To compute the background evolution, we start from the $f(R)$ field equation for a FLRW metric
\begin{eqnarray}\label{eq:fofrback}
        H^2 (1 + f_R + R' f_{RR}) - \frac{Rf_R-f}{6} = m^2 e^{-3y}, \label{einstein_eq}
\end{eqnarray}
where $^\prime \equiv d/dy$ ($y=\ln a$), $H \equiv \dot{a}/a$ is the Hubble parameter (a dot denotes the derivative with respect to cosmic time), which is related to R by
\begin{equation}
R=12 H^2 + 6 H H^\prime. \label{R}
\end{equation}
To solve these equations, we introduce the new variables
\begin{align}
x_1 (y) &= \frac{H^2}{m^2} - e^{-3y} - d,\\
x_2 (y) &= \frac{R}{m^2} - 3 e^{-3y} - 12\left(d + x_1\right).
\end{align}
With these definitions we obtain
\begin{align}
x'_1 (y) &= \frac{x_2}{3},\\
x'_2 (y) &= \frac{R'}{m^2} + 9 e^{-3y} - 4 x_2,
\end{align}
where $R'$ is given by Eq.~(\ref{einstein_eq}). It is straightforward to verify that, as defined,  $x_1$ and $x_2$ are always zero during the $\Lambda$CDM phase. We here focus on the three cases summarized in Table~\ref{model_tab}.

Furthermore, in terms of $x_1(y)$ and $x_2(y)$,  the effective dark energy equation of state ($w_{\rm DE}$) is given by,\begin{equation}
w_{\rm DE} = -1 - \frac{1}{ 9} \frac{x_2}{x_1 + d}.
\end{equation}
For the considered models, the evolution of $w_{\rm DE} $ as a function of the redshift $z$ is shown in Fig.~\ref{steq}.

\begin{table}
\begin{center}
\setlength{\tabcolsep}{.05\columnwidth}
\begin{tabular}{ccc}
        \hline
        $\tilde{\Omega}_{m0}$ & $n$ & $\alpha$ \\
        \hline \hline
    0.267 \qquad & 2 \qquad & 1.05 \\
    0.267 \qquad & 2 \qquad & 1.18 \\
    0.267 \qquad & 2 \qquad & 1.5 \\
        \hline
\end{tabular}
\caption{Overview of the model parameters for $\gamma$ gravity.}\label{model_tab}
\end{center}
\end{table}

\begin{figure}
\includegraphics[width=1.0\columnwidth]{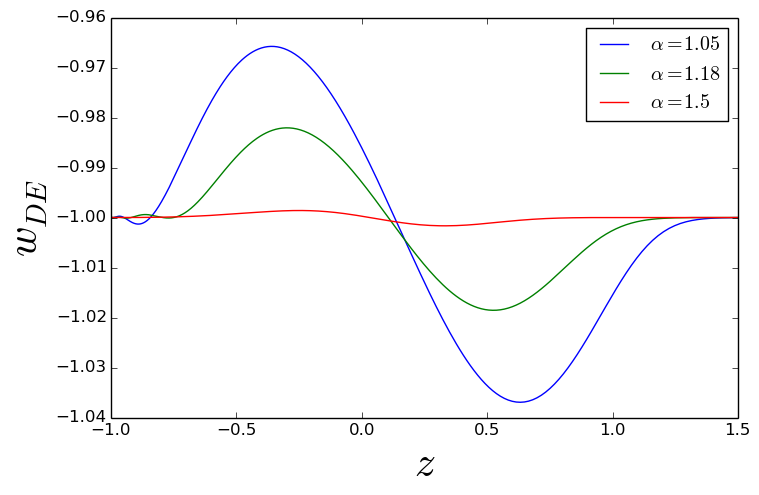}
\caption{Effective equation-of-state parameter $w_{\rm DE}$ as a function of redshift $z$ for the parameters given in Table \ref{model_tab}. The strongest deviation from $-1$ is lower than $4\%$.}\label{steq}
\end{figure}

For a general $f(R)$ model the differential equation for the matter density contrast ($\delta_m$) in the  linear regime for subhorizon scales is given by \citep{Pogosian:2007sw,Zhang:2005vt,delaCruzDombriz:2008cp}
\begin{align}
\delta''_m + \left(2+\frac{H'}{H}\right)\delta'_m - \frac{1-2Q}{2-3Q} \frac{3H_0^2 \tilde{\Omega}_{m0}}{H^2(1+f_R)}e^{-3y}\delta_m = 0, \label{linear}
\end{align}
where
\begin{equation}
        Q(k,y) = -\frac{2f_{RR}c^2 k^2}{(1+f_R)e^{2y}}
.\end{equation}

In GR, $f_{RR} = Q = 0,$ and there is no scale dependence for the density contrast in the linear regime. For $\Lambda$CDM the growing mode can be expressed in terms of hypergeometric function ${}_2 F_1$ as \citep{Silveira:1994yq}

\begin{equation}
        \delta_+ \propto e^{-y} {}_2 F_1 \left[ \frac{1}{3} , 1 , \frac{11}{6} , -e^{3y}d \right]. \label{growth_lcdm}
\end{equation}

We solved Eq. (\ref{linear}) numerically and obtained the growing mode for the $\gamma$ gravity. By using (\ref{growth_lcdm}), we then obtained the fractional change in the matter power spectrum $P(k)$ relative to $\Lambda$CDM. Figure \ref{linear_fig} shows $\Delta P_k/P_{\Lambda}$ at $z = 0$ for the three choices of parameters shown in Table~\ref{model_tab}.
        
\begin{figure}
\label{fig:linearpofk}
\includegraphics[width=1.0\columnwidth]{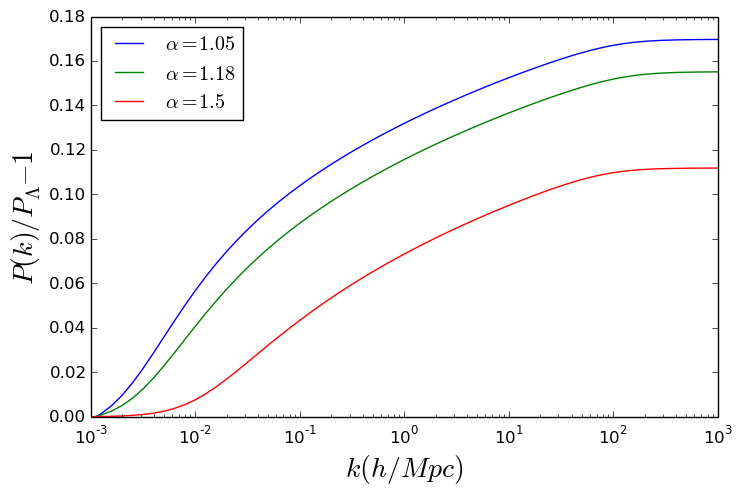}
\caption{Fractional difference in the matter power spectrum with respect to $\Lambda$CDM for different values of $n$ and $\alpha$, as indicated in Table \ref{model_tab}. These results are used to compare with the nonlinear power spectrum from our numerical simulations.}\label{linear_fig}
\end{figure}


\section{{\it N}-body equations}\label{nbody}

$f(R)$ models are equivalent to a scalar-tensor theory \citep{2008PhRvD..78j4021B}, where the first derivative of the $f(R)$ function, $f_R$. This field propagates according the equation

\begin{equation}
        \square f_R = \frac{\partial V_{\rm eff}}{\partial f_R} = \frac{(1-f_R)R + 2f + \kappa^2 T}{3},
\end{equation}
where $\kappa^2 = 8\pi G/c^4$ and $T$ is the trace of energy-momentum tensor, $T = g_{\mu\nu}T^{\mu\nu}$. In the quasi-static limit (see, e.g., \cite{2014PhRvD..89b3521N,2015JCAP...02..034B,Llinares:2013jua,Llinares:2013qbh}) this equation becomes
\begin{align}
\frac{1}{a^2}\nabla^2 f_R = \frac{R - R(a)}{3} - m^2 a^{-3} \delta_m,
\end{align}
where
\begin{align}
\frac{R(a)}{m^2} = 3(a^{-3} + 4d) + \Delta_{R}(a),
\end{align}
and $\Delta_{R}(a) = x_2(a) + 12 x_1(a)$. The Ricci scalar $R$ in function of $f_R$ is given by inverting Eq.~(\ref{fReq})
\begin{align}
R = R_*\log\left(\frac{\alpha}{|f_R|}\right)^{1/n}.
\end{align}
The geodesic equation, needed to update the particle positions, reads
\begin{align}
\ddot{x} + 2H\dot{x}  = -\frac{1}{a^2}\nabla\left(\Phi - \frac{f_R}{2}\right),
\end{align}
where $\Phi$ is the newtonian potential, which the dynamics is given by the Poisson equation
\begin{align}
\nabla^2\Phi = \frac{3m^2}{2}\frac{\delta_m}{a}.
\end{align}
When implementing these equations in the {\it N}-body code, we need to rewrite them in code-units given by
\begin{align}
\tilde{x} &= x/B_0,~~~~ \tilde{\Phi} = \frac{\Phi a^2}{(H_0B_0)^2},\nonumber\\
d\tilde{t} &= \frac{H_0dt}{a^2},~~~~\nabla_{\rm code} = B_0.\nabla.
\end{align}
Here $B_0$ is the size of the simulation box. In terms of $\tilde{f}_{R} = -a^2f_R,$ the evolution equations becomes
\begin{align}\label{eq:alleq}
&\frac{d^2\tilde{x}}{d\tilde{t}^2} = -\nabla_{\rm code} \tilde{\Phi} - \frac{1}{2(B_0H_0)^2}\nabla_{\rm code} \tilde{f}_{R}, \\
&\nabla_{\rm code}^2\tilde{\Phi} = \frac{3}{2}\Omega_{m0} a \delta_m, \\
&\nabla_{\rm code}^2 \tilde{f}_R = \Omega_{m0} (H_0B_0)^2 a^4 \times \\
&\times \left\{-\frac{R_*}{3m^2} \log\left(\frac{\alpha a^2}{\tilde{f}_R}\right)^{1/n} + \left[ a^{-3} + 4d + \frac{\Delta_R (a)}{3}\right] + a^{-3}\delta_m\right\}.\nonumber
\end{align}
These are the only equations we need to implement and solve in the {\it N}-body code.

For comparison we also need the linearized field equation. Simulations with this equation compared to the full $f_R$ equation is a good measure of the amount of screening that takes place in the model. The linearized $f_R$ equation is simply
\begin{align}
\frac{1}{a^2}\nabla^2\delta f_R = m_{\phi}^2(a)\delta f_R - m^2 a^{-3} \delta_m,
\end{align}
where $\delta f_R = f_R - f_R(a)$ and $m_{\phi}^2(a) = \frac{1}{3f_{RR}(a)}$. In code units, taking $u = -\frac{\delta f_R a^2}{2(H_0 B_0)^2}$, we obtain\begin{align}
\nabla_{\rm code}^2 u = [m_{\phi}(a)aB_0]^2 u + \delta_m \frac{\Omega_{m0} a}{2},
\end{align}
and the geodesic equation becomes
\begin{align}
\frac{d^2\tilde{x}}{d\tilde{t}^2} = -\nabla_{\rm code} \tilde{\Phi} - \nabla_{\rm code} u.
\end{align}
We have
\begin{align}
m_{\phi}^2(a)a^2B_0^2 = \frac{a^2(H_0 B_0)^2}{3\alpha n}\frac{R(a)}{H_0^2} \left[\frac{R_*}{R(a)}\right]^ne^{\left[R(a)/R_*\right]^n}.
\end{align}


\section{Implementation in the {\tt{ISIS}} code}\label{isis}

Implementing scalar-tensor theories of gravity in {\it N}-body code is rather straightforward because the scalar-tensor theories all contribute as a fifth force and because {\tt{RAMSES}}, which {\tt{ISIS}} is based on, has been widely used, thoroughly tested, and optimized. In this section we describe how the equations we need to solve are implemented in {\tt{ISIS}}. For more details see \citep{Llinares:2013jza}. 

To solve for $f_R$ directly is not numerically stable since the solution can potentially vary over several orders of magnitude when going from deep voids to massive clusters in our simulation. We therefore introduce a field redefinition $|f_R| = A(u) \to \nabla |f_R| = b(u)\nabla u$ where $b(u)=dA(u)/du$. The general field equation for $f_R$ discretized on a grid with the field-redefinition $\tilde{f}_R \equiv -a^2f_R = A(u)$, where $u$ is the field we solve for, can be written as
\begin{align}
&\mathcal{L}(u_{i,j,k}) = \nabla_{\rm code}\cdot [b(u)\nabla_{\rm code} u]_{i,j,k} + \Omega_{m0} (H_0B_0)^2 \times\\
&\times\left\{\frac{a^4 R_*}{3m^2} \log\left[\frac{\alpha a^2}{A(u_{i,j,k})}\right]^{1/n} - a(\delta_m)_{i,j,k} - a - a^4 \left[ 4d + \frac{\Delta_R (a)}{3} \right] \right\},\nonumber
\end{align}
where
\begin{align}
&\nabla_{\rm code}\cdot [b(u)\nabla_{\rm code} u]_{i,j,k} =\\ &\frac{b_{i+1/2,j,k}(u_{i+1,j,k}-u_{i,j,k}) - b_{i-1/2,j,k}(u_{i,j,k}-u_{i-1,j,k})}{h^2}+\nonumber\\
&+\frac{b_{i,j+1/2,k}(u_{i,j+1,k}-u_{i,j,k}) - b_{i,j-1/2,k}(u_{i,j,k}-u_{i,j-1,k})}{h^2}+\nonumber\\
&+\frac{b_{i,j,k+1/2}(u_{i,j,k+1}-u_{i,j,k}) - b_{i,j,k-1/2}(u_{i,j,k}-u_{i,j,k-1})}{h^2},\nonumber
\end{align}
where $h$ is the grid spacing and $b_{i\pm 1/2,j,k} \equiv \frac{1}{2}(b(u_{i \pm 1,j,k}) + b(u_{i,j,k}))$ and where we have defined $b(u) \equiv \frac{dA(u)}{du}$.

The equations are solved using Newton-Gauss-Seidel relaxation (with multigrid acceleration). The method consists of going through the grid and updating the solution using
\begin{align}
u_{i,j,k}^{\rm new} = u_{i,j,k} - \frac{\mathcal{L}(u_{i,j,k})}{\partial \mathcal{L}(u_{i,j,k})/\partial u_{i,j,k}}.
\end{align}
For this we also need $\mathcal{\partial L}(u_{i,j,k})/\partial u_{i,j,k}$ , which is given by
\begin{align}
&\frac{\partial\mathcal{L}(u_{i,j,k})}{\partial u_{i,j,k}} = \frac{\partial \nabla_{\rm code}\cdot [b(u)\nabla_{\rm code} u]_{i,j,k}}{\partial u_{i,j,k}} + \\ - &\Omega_{m0} (H_0B_0)^2\left\{\frac{a^4 R_*}{3m^2}\frac{b(u_{i,j,k})}{A(u_{i,j,k})} \log\left[\frac{\alpha}{A(u_{i,j,k})}\right]^{1/n-1}\right\},\nonumber
\end{align}
where
\begin{align}
&\frac{\partial \nabla_{\rm code}\cdot [b(u)\nabla_{\rm code} u]_{i,j,k}}{\partial u_{i,j,k}} = \\
& -\frac{b_{i+1/2,j,k} + b_{i-1/2,j,k}}{h^2} - \frac{b_{i,j+1/2,k} - b_{i,j-1/2,k}}{h^2} +  \nonumber\\
&-\frac{b_{i,j,k+1/2} + b_{i,j,k-1/2}}{h^2} + \frac{1}{2}c_{i,j,k}\left[\frac{u_{i+1,j,k} + u_{i-1,j,k} - 2u_{i,j,k}}{h^2} + \right.\nonumber\\
&\left. + \frac{u_{i,j+1,k} + u_{i,j-1,k} - 2u_{i,j,k}}{h^2} + \frac{u_{i,j,k+1} + u_{i,j,k-1} -2u_{i,j,k}}{h^2}\right],\nonumber
\end{align}
and $c_{i,j,k} = c(u_{i,j,k}),$ where $c(u) = \frac{db(u)}{du}$. Some problems arise related to how boundary conditions are handled on the refined grids in the code. This is discussed in \cite{Llinares:2013jza,Li:2011vk}.

Our solver needs a starting guess, and for this we use the cosmological background solution, that is, the solution we would expect when there are no matter sources,
\begin{align}
|f_R(a)| = \alpha e^{-\left[R(a)/R_*\right]^n} \to \overline{u} = A^{-1}(|f_R(a)|).
\end{align}
This is only needed when the simulation is stared as otherwise we can use the old solution as our guess. For $\gamma$ gravity our choice for $A$ and the related expressions for $b$ are
\begin{align}
&A(u) \equiv \tilde{f_R}(u) = \alpha a^2 e^{-e^u},\\
&b(u) = \frac{dA(u)}{du} = -\alpha a^2 e^{-e^u}e^u,\\
&c(u) = \frac{db(u)}{du} = \alpha a^2 e^{-e^u}(e^u-1)e^u,
\end{align}
and the background value of $u$ is
\begin{align}
\overline{u} = A^{-1}(\tilde{f_R}(a)) = n\log\left[\frac{a^{-3} + 4d + \Delta_R (a) / 3}{R_*/(3H_0^2)}\right],
\end{align}
in terms of $u$ we have
\begin{align}
&\mathcal{L}(u) = \nabla[b(u)\nabla u] + \Omega_{m0} (H_0B_0)^2 \times \\ &\times\left[\frac{a^4 R_*}{3m^2} e^{u/n} - a\delta_m - a - a^3 \left(4d + \frac{\Delta_R (a)}{3} \right)\right], \nonumber\\
&\frac{\partial}{\partial u} \mathcal{L}(u) = \frac{\partial}{\partial u}\nabla[b(u)\nabla u] + \Omega_{m0} (H_0 B_0)^2\left( \frac{a^4R_*}{3m^2} \frac{e^{u/n}}{n} \right).
\end{align}
When the fifth force is implemented,  we simply replace it with an effective force $F_{\rm eff}$ that includes the effects of modified gravity wherever the code normally
works with the gravitational force, $F_{N}$,
\begin{equation}
F_{\rm eff} = F_N + F_{\phi}.
\end{equation}
The expression for the fifth force in code-units is given in Eq.~(\ref{eq:alleq}), and we use the same $\text{five}$-point stencil as {\tt{RAMSES}} uses to compute the gravitational force $\nabla\tilde{\Phi}$ to compute the fifth force $\nabla\tilde{f}_R$.

\section{Tests of the {\it N}-body solver} \label{test}

\begin{figure*}
\includegraphics[width=2.0\columnwidth]{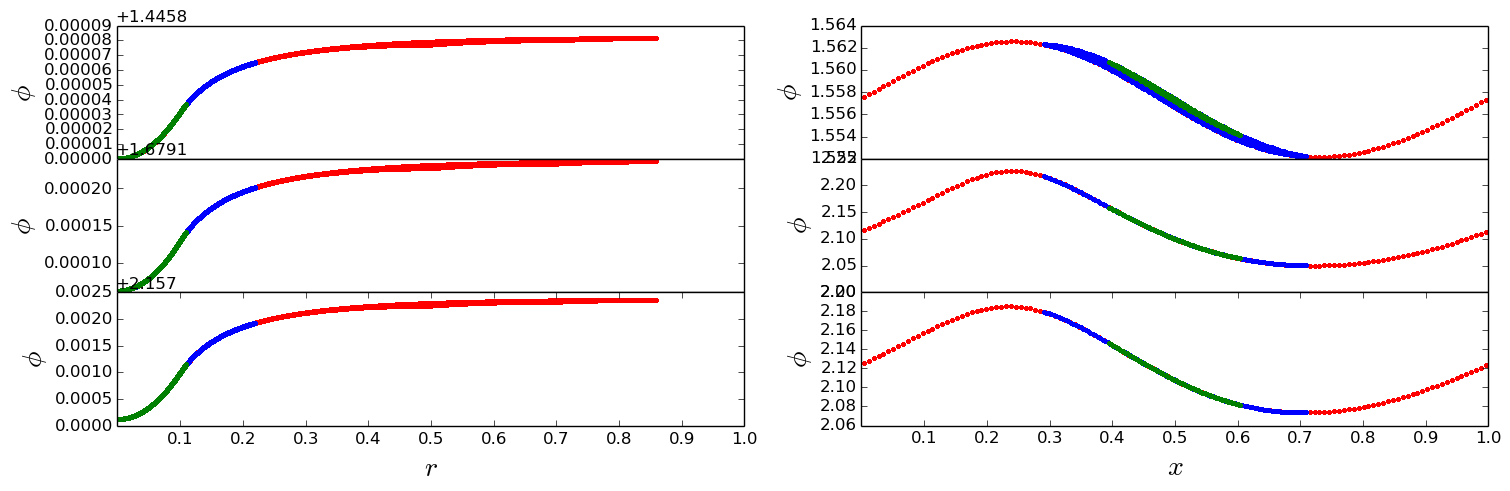}
\caption{Scalar field from tests. Different colors depict the different refinement levels. The left panel shows the field from a spherical density distribution, the right side shows a $1$D sine field obtained in the second test.}\label{tests}
\end{figure*}

To verify that the solver is implemented correctly, we tested it several times. We present some of these tests in this section. 
The density is provided to the code through a distribution of particles. The density estimation (CIC) and refinement criteria are the same as those used for the cosmological simulations. We also have the option to set the density field analytically in the code. To test that the treatment of the boundary of the refinement is correct, we included two levels of refinement. The model parameters used for the tests are the same as mentioned previously in Table~\ref{model_tab}.

The first test case corresponds to a sphere of radius $R$ of constant density located in the center of the box and embedded in a uniform background:
\begin{equation}
        \rho(r) = \left\{ \begin{array}{c}\displaystyle\frac{(1+\delta)\overline{\rho}}{1+\frac{4\pi}{3}\delta\left(\frac{R}{B_0}\right)^3}, \qquad r < R \\ \\
\displaystyle\frac{\overline{\rho}}{1+\frac{4\pi}{3}\delta\left(\frac{R}{B_0}\right)^3}, \qquad r > R
\end{array} \right. ,
\end{equation}
where $\delta = \frac{\rho_{\rm in}}{\rho_{\rm out}}-1$ characterizes the density contrast between the inside and outside of the sphere,  $\bar{\rho}$ is the mean density, $R$ is the radius of the sphere, and $B_0$ the size of the box. The value of $\delta$ chosen for the test is $5000$. For the $f(R)$ test we used $R = 25$ Mpc$/$h and $B_0 = 250$ Mpc$/h$. A spherical symmetric configuration is effectively one-dimensional, therefore the field equation reduces to an ODE, which we solved by using Mathematica and used this to compare with.

For the second test we analytically computed a density field $\rho(x,y,z) \equiv \rho(x)$ (i.e., a $1$D configuration), using the field equation, so that the solution is given by a sine: $u \propto 2+\sin(2\pi x)$. 

Figure~\ref{tests} shows the result of both these tests and the different colors depict the different refinement levels. We see that the curves are smooth when going from one level to another, which demonstrates that boundary conditions are handled properly. The tests were performed using the serial version of the code, and both tests give the expected results, which demonstrates that the code works properly.

\section{Simulations}

The $\gamma$ gravity simulations were run using $512^3$ dark matter particles with a box size of $B_0 = 250$ Mpc$/h,$ and we refined cells whenever it had more than $\text{eight}$ particles in it. The highest refinement level reached in our simulation was eight. The background cosmology used for the simulation was computed using Eq.~(\ref{eq:fofrback}) with $h=0.71$, $\Omega_{\Lambda} = 0.733$ and $\Omega_{m0} = 0.267$.

The model parameters were presented in Sect.~\ref{model}, with some plots of various background quantities. We also compared these simulations with simulations of the Hu-Sawicki $f(R)$ model from \cite{Llinares:2013jza}.

\section{Results}\label{results}

\subsection{Power spectrum}

The nonlinear matter power spectrum is an important observable and could be used to distinguish among different models of structure formation. As we showed above, $\gamma$ gravity can have a strong effect on the growth rate of the linear perturbations. We expect these signatures to be detectable in the nonlinear matter power spectrum.

To compute the power spectrum we used a public code, {\tt POWMES} \citep{2011ascl.soft10017C}, which uses folding methods to compute the power spectrum. 

Figure~\ref{powerspec} displays the 
difference of the matter power spectrum with respect to $\Lambda$CDM, defined as $\Delta P/P_{\Lambda\rm CDM} \equiv P(k)/P_{\Lambda CDM}(k) - 1$ for $\gamma$ gravity together with the corresponding predictions from linear perturbations theory and the Hu-Sawicki $f(R)$ model for comparison. We focus on the present-day epoch, which corresponds to $z = 0$.

Figure~\ref{powerspec} shows that the differences from $\Lambda$CDM are lower than $5-10\%$ for all of our runs in the range $0.05 h\text{Mpc}^{-1} \lesssim k \lesssim 0.1 h\text{Mpc}^{-1}$) and in agreement with the linear perturbation result seen in Fig.~(\ref{fig:linearpofk}). The deviation from $\Lambda$CDM approaches zero for larger scales. This is because the range of fifth force is smaller than the horizon, therefore the modifications of gravity are not felt on the largest scales. On smaller, nonlinear scales, the full effect of the fifth force acts, and we see larger deviations. However, $\Delta P/P_{\Lambda \rm CDM}$   continues to grow with $k$ in $\gamma$ gravity, in contrast to the Hu-Sawicky model, where the chameleon screening mechanism is in play on these scales,
which reduces the power enhancement. This is an indication that the screening mechanism does not work very well in our simulation. We discuss this in more detail in Sect. \ref{sect:ffandscreen}.

\begin{figure}
\includegraphics[width=1.0\columnwidth]{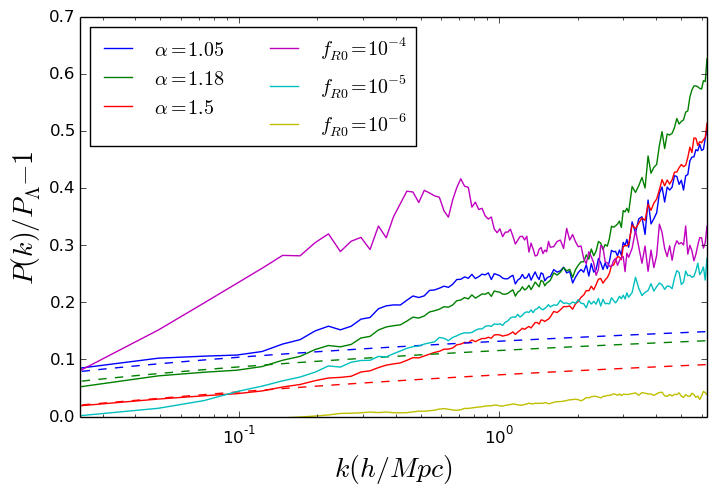}
\caption{Fractional deviation of the dark matter power spectrum with respect to the $\Lambda$CDM model. The linear predictions of $\gamma$ gravity are represented by dashed lines. For comparison we also show the results from simulations with the Hu-Sawicky $f(R)$ model with $|f_{R0}|=\{10^{-4},10^{-5},10^{-6}\}$ from \cite{Llinares:2013jza}. }\label{powerspec}
\end{figure}

\subsection{Halo mass function}

The halo mass function, the number density of halos with a given mass, is a useful tool for investigating the efficiency of a model in forming halos of different masses. To locate halos in the simulation outputs, we used the {\tt AHF} (Amiga Halo Finder \citep{Knollmann:2009pb}). We determined the halo mass function by binning halos in logarithmic mass intervals.

In Fig.~\ref{massfunc} we show the fractional difference with respect to $\Lambda$CDM of halo mass function computed from our simulations. Our measurement of the halo mass function itself is limited by statistics and to a lesser extent, by the resolution in the high and low mass end. However, we can reduce the impact of these two effects by considering the relative difference between the halo mass functions measured in modified gravity and $\Lambda$CDM simulations with the same initial condition.

Figure~\ref{massfunc} shows that we find an excess of halos in the range $10^{13} \lesssim M/M_{\odot} \lesssim 10^{15}$ probed by our simulation in $\gamma$ gravity, and the signatures are very similar to the $|f_{R0}| \gtrsim 10^{-5}$ Hu-Sawicki model. For the most massive halos ($M/M_{\odot} \gtrsim 10^{15}$) the effect of screening is much more pronounced in Hu-Sawicki models,
and the mass-function approaches $\Lambda$CDM. This does not occur for $\gamma$ gravity and is again an indication that the chameleon screening mechanism does not work very efficiently in our simulations.

For $\gamma$ gravity, the number of halos increases significantly, especially at the high mass end, by up to $40-100\%$ for cluster-sized halos. For Hu-Sawicki models, when the value of the $f_R$ field becomes comparable to the cosmological potential wells, the chameleon effect starts to operate. This can be seen in the mass function, where deviations from $\Lambda$CDM approach zero in the high-
mass end for models with $|f_{R0}| = 10^{-5}$ and $|f_{R0}| = 10^{-6}$.

The large deviations we find $\Delta n/n_{\Lambda\rm CDM} \sim 0.5-1$ in the high-mass end are probably already ruled out by present cluster counts \citep{2015PhRvD..92d4009C,2012PhRvD..85l3513M}. This does not rule out the model per se, but requires choices of parameters where $|f_{R0}|$ is much smaller today, leading to no signatures in the background evolution (i.e., in the Hubble factor) of the Universe.

\begin{figure}
\includegraphics[width=1.0\columnwidth]{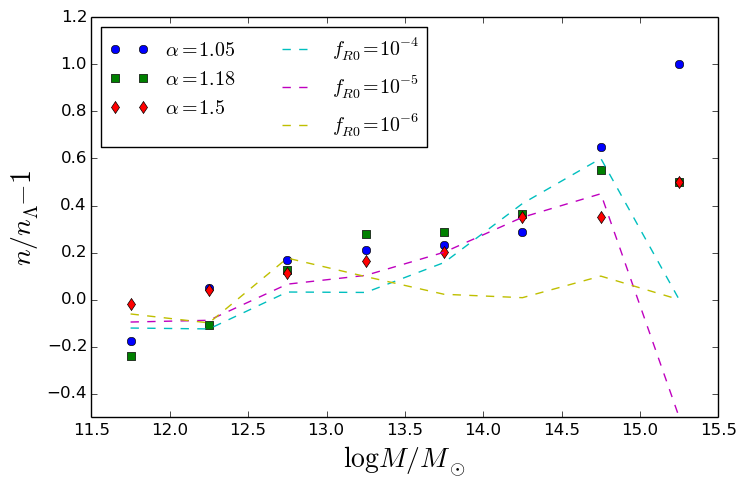}
  \caption{Fractional difference of the halo mass function for $\gamma$ gravity (data points) with respect to the $\Lambda$CDM model. For comparison we also show the results from simulations with the Hu-Sawicky $f(R)$ model with $|f_{R0}|=\{10^{-4},10^{-5},10^{-6}\}$ (dashed lines) from \cite{Llinares:2013jza}.}\label{massfunc}
\end{figure}

\subsection{Halo profiles}

\begin{figure*}
\includegraphics[width=2.0\columnwidth]{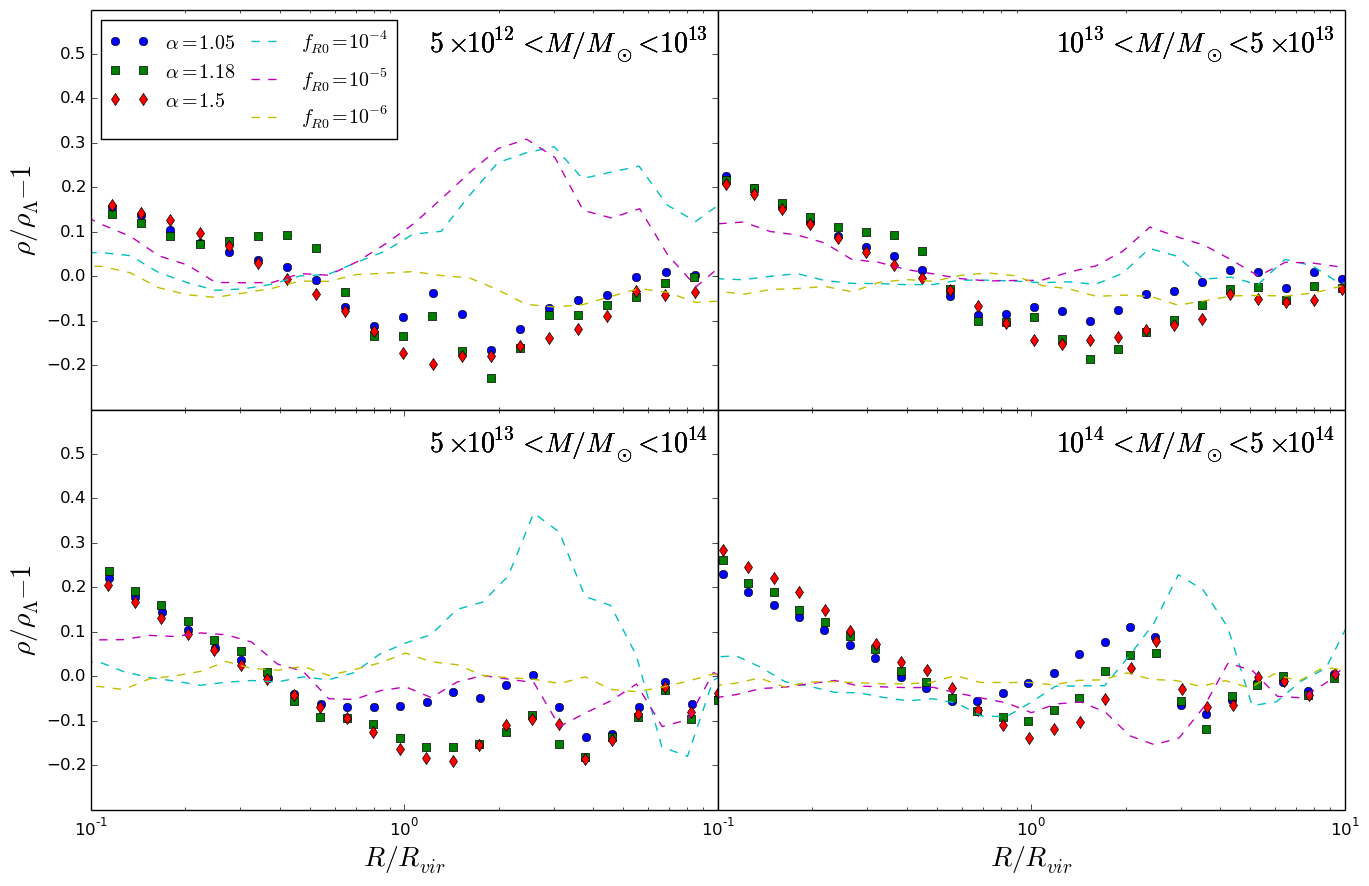}
\caption{Fractional difference in the halo density profiles with respect to $\Lambda$CDM for four different mass bins. For comparison we also show the results from simulations with the Hu-Sawicky $f(R)$ model with $|f_{R0}|=\{10^{-4},10^{-5},10^{-6}\}$ (dashed lines) from \cite{Llinares:2013jza}.}\label{dens_prof}
\end{figure*}

We also studied how modifications of gravity change the density and velocity profiles of dark matter halos. We focused on halos in the mass ranges $0.5-1 \times 10^{13}$Mpc$/h$, $1-5 \times 10^{13}$Mpc$/h$, $0.5-1 \times 10^{14}$Mpc$/h,$ and $1-5 \times 10^{14}$Mpc$/h$. The density profiles were calculated by binning dark matter particles in annular bins for each halo. Our calculated density profiles are averages of all density profiles of the proper size, ranging from 10\% of the virialization radius, $r = 0.1 R_{\rm vir}$, to ten times the virialization radius, $r = 10 R_{\rm vir}$. This range was chosen to properly include all behaviors of the fifth force on the dark matter halos while also avoiding the inner regions of the halos, where the resolution of our simulations is not sufficient.

Figure~\ref{dens_prof} shows the fractional difference with respect to $\Lambda$CDM in the density profiles. We first note that the inner regions ($R < R_{\rm vir}$) of halos for $\gamma$ gravity are significantly denser than in $\Lambda$CDM. This difference is compensated for in outer regions ($R>R_{\rm vir}$). Moreover, the profiles between the different model parameters do not differ appreciably from each other, and this pattern repeats for all ranges. However, the density profiles for Hu-Sawicki models in
general show stronger clustering in the low-density regions in the outskirts of halos than in the inner regions.

In the velocities profiles, shown in Fig.~\ref{vel_prof}, we expect the fifth force to increase the velocity dispersion. This effect is very similar for both models, except for most massive halos, for which the Hu-Sawicki models are more screened, causing a substantial decrease in comparison to $\gamma$ gravity.
For the Hu-Sawicki model the velocities are boosted by $\sim 20\%$ in the $|f_{R0}| = 10^{-4}$ case and by $5\%$ in the $|f_{R0}| = 10^{-6}$ case for all the three halo mass ranges analyzed. Only for the $|f_{R0}| = 10^{-5}$ parameters a mixed behavior is seen, that is, for the lower two halo mass ranges the boost is $\sim 20\%$, but for the heaviest halos there is no deviation from $\Lambda$CDM in the inner parts and $\sim 15\%$ higher velocities are found in the outer parts. For $\gamma$ gravity, the difference between the models we simulated is more expressive for less massive halos ($5\times10^{12} < M/M_{\odot} < 10^{13}$ and $10^{13} < M/M_{\odot} < 5\times 10^{13}$), for most massive halos ($10^{14} < M/M_{\odot} < 5\times 10^{14}$) only the case $\alpha = 1.5$ is distinguishable from the others.

\begin{figure*}
\includegraphics[width=2.0\columnwidth]{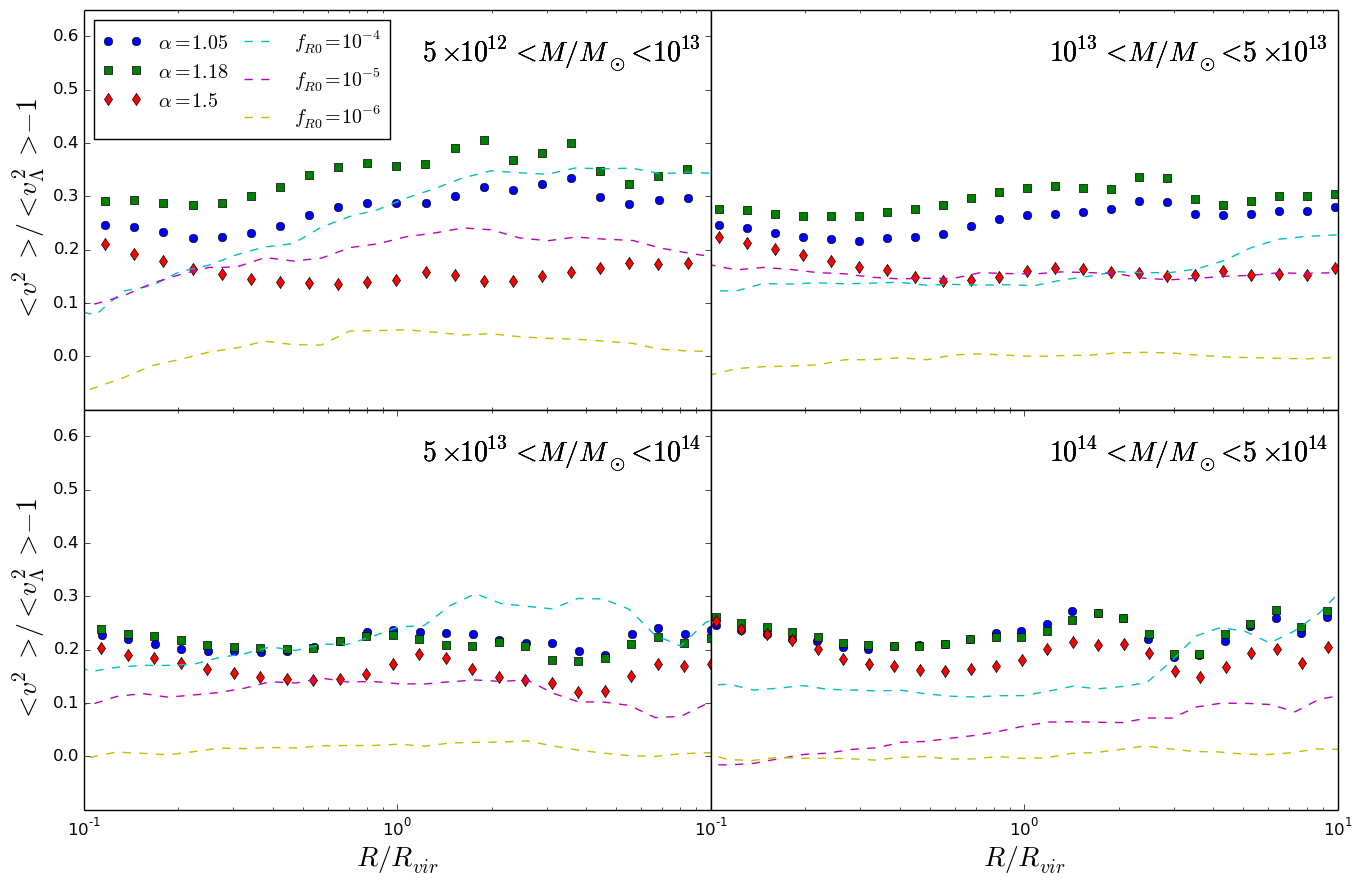}
\caption{Fractional difference in the velocity dispersion profiles with respect to $\Lambda$CDM for four different mass bins. For comparison we also show the results from simulations with the Hu-Sawicky $f(R)$ model with $|f_{R0}|=\{10^{-4},10^{-5},10^{-6}\}$ (dashed lines) from \cite{Llinares:2013jza}.}\label{vel_prof}
\end{figure*}

\subsection{Fifth force and screening}\label{sect:ffandscreen}

General relativity is very well tested on very small scales, especially inside the solar system. To ensure that $F_{\phi}$ is not relevant at these scales, we need a screening mechanism to suppress the fifth force at small-scale, very high curvature regimes.

When $\gamma$ gravity was proposed in \cite{O'Dwyer:2013mza}, the authors explored the compatibility between the model and the solar system experiments using the chameleon mechanism for screening, as any other $f(R)$ \citep{2008PhRvD..78j4021B}. 

As we showed in the results above, the screening mechanism does not seem to be working very efficiently for the models simulated here. To check how much the fifth force is screened in our simulations, we compared the magnitude of the fifth and Newtonian force on the particles in our simulation box, as in \cite{Davis:2011pj}.

Figure~\ref{fifth} shows a scatter plot for this comparison at redshift $z=0$. The dispersion for small $F_N < 0.01$ is expected (numerical scatter), here the forces are tiny, so the scatter here has a very weak effect on the simulation. The important part in this figure is the behavior for large $F_N$. If we have screening, then we should see $F_\phi \ll F_N/3$ in high-density regions (which corresponds to high values of $F_N$). The result we find shows that the force ratio is roughly $\text{one third}$ - which is the linear prediction - everywhere, meaning that there is very little screening present in our simulations.

To understand this better, we revisit the screening conditions. Considering a spherical symmetric body with constant density $\rho_c$ embedded in the background of constant density $\rho_b$ , the solutions to the field equation (see e.g. \cite{2012PhRvD..86d4015B}) mean that the fifth force is given by
$$F_{\phi} \sim \frac{1}{3}\frac{\Delta R}{R} \frac{GM}{r^2} e^{-m_b r},$$
where
$$\frac{\Delta R}{R} = \frac{|f_{Rc} - f_{Rb}|}{2\Phi_N},$$
is the so-called screening factor (also called the thin-shell) and $f_{Rb}$ ($f_{Rc}$) is the value of $f_R$ in the background (inside the body), where $\rho=\rho_b$ ($\rho_c$). If $\frac{\Delta R}{R} \ll 1,$ the fifth force is screened and we recover General Relativity. If $\frac{\Delta R}{R} \gtrsim 1,$ we instead find that $F_\phi \propto \frac{1}{3}F_N$ and gravity is significantly modified.

When the field is located in the minimum of its effective potential in the background\footnote{Background here refers to the surroundings for the body in question, not necessarily the cosmological background.} , we can calculate the screening factor analytically. Assuming this, we find that for the $\gamma$ gravity models considered here, Earth and Sun are almost completely screened and pass local gravity constraints assuming the galaxy is screened.

However, for the parameter values considered in this paper, the screening condition gives that the galaxy is not screened which again implies that the Earth and the sun is not screened either\footnote{This is only true for the parameters considered in this paper. If $|f_{R0}| \lesssim 10^{-5}$ , then the galaxy, Earth, and our Sun are screened.}. The only caveat to this is that screening might have survived from earlier times. At early redshift $|\overline{f}_{R}| \ll 1$ and almost all objects are screened. Then in the cosmological evolution $|\overline{f}_{R}|$ very quickly evolves to high values $|\overline{f}_{R}| = O(0.01-0.1)$. The field inside our galaxy might have been trapped, ensuring screening. This is not expected, but to rigorously rule out (or confirm) this possibility, we would need simulations beyond the quasi-static approximation that we assumed in our simulations. This is beyond the scope of this paper.

\begin{figure}
\includegraphics[width=1.0\columnwidth]{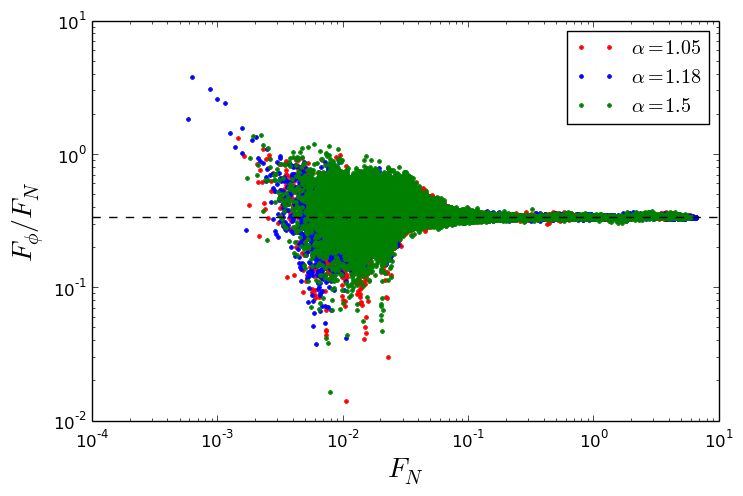}
\caption{Comparison between the magnitudes of Newtonian force ($F_N$) and fifth force ($F_{\phi}$), the dashed line indicates $F_N/F_{\phi} = 1/3$. The forces are in units of $H_0/c^2$.}\label{fifth}
\end{figure}

\section{Summary and conclusions} \label{conc}

We have investigated the nonlinear evolution in $\gamma$ gravity, a $f(R)$ theory of gravity that is a viable alternative to $\Lambda$CDM. The models we investigated use a screening mechanism to suppress the deviations from General Relativity at small (solar system) and large cosmological scales. Specifically, this is the chameleon screening mechanism. As a result of  this screening mechanism, the strongest signatures in these models are expected to occur at the nonlinear regime of structure formation. Therefore, to unveil the imprints of such theories at astrophysical scales, we ran several cosmological {\it N}-body simulations. We compared models with $\Lambda$CDM and the Hu-Sawicki model and showed that several astrophysical observables (halo mass function, density profiles, and power spectra) show the signatures of the model
significantly strongly.

For the matter power spectrum we found a small deviation, lower than 10\%, on large scales ($k \lesssim 10^{-1}~h$/Mpc), which
is consistent with the predictions of linear perturbation theory. For small scales ($k \gtrsim 10^{-1}~h/$Mpc), on the other hand, we found a strong increase in power, the largest deviation ($\alpha = 1.18$) reaches $\sim 60\%,$ and in the other cases ($\alpha = 1.05$ and $\alpha = 1.5$) it reaches $\sim 50\%$. This is quite different from the results found in the Hu-Sawicky model, where the screening of the fifth force is much more active in suppressing the enhancement of power on small scales, the largest deviation is $\sim 30\%$, much lower than for $\gamma$ gravity.

For the halo mass function we found that all of our runs gave very similar result in the mass range $ 11.5 < \log (M/M_{\odot}) < 14.7$, and these results are very similar to what is found in the Hu-Sawicky model for $|f_{R0}|\geq 10^{-5}$; a $10-60\%$ increase in the halo abundance. However, for $\gamma$ gravity we found an enhancement of $40-100\%$ in the abundance for $\log (M/M_{\odot}) > 14.5,$ which can be compared to the Hu-Sawicki
model, where the halo mass function approaches that of $\Lambda$CDM.

For halo density profiles we found that the different runs with $\gamma$ gravity were not distinguished well in any mass range; in all situations we saw that the inner regions ($R < R_{\rm vir}$) of halos for $\gamma$ gravity are significantly thicker than the halos of $\Lambda$CDM, around $10\%$ for less massive halos, $5\times 10^{12} < M/M_{\odot} < 10^{13}$, increasing with the halo mass until they reached $\sim 30\%$ for the most massive halos, $10^{14} < M/M_{\odot} < 5\times 10^{14}$. This difference is compensated for in the outer regions ($R>R_{\rm vir}$). For the Hu-Sawicki model the same effect acts for less massive halos $5\times 10^{12} < M/M_{\odot} < 10^{13}$, but it changes for more massive halos, $5\times 10^{13} < M/M_{\odot} < 10^{14}$ and $10^{14} < M/M_{\odot} < 5\times 10^{14}$ when the screening mechanism suppresses the fifth force. This then leads to a suppression in the accretion of mass concentration in the inner regions (compared to the non-screened case).

Finally, we computed the velocity profiles, and as we expected, the velocities are significantly enhanced in $\gamma$ gravity compared to $\Lambda$CDM. The other cases are comparable in all mass ranges, but with some important signatures. The difference between the different $\gamma$ gravity models is larger for low-mass halos, $5\times 10^{12} < M/M_{\odot} < 10^{13}$, and we can distinguish them from each other; the differences reach $\sim 30\%$ close to the boundary region, $R\sim R_{\rm vir}$. This difference decreases with the mass of the halos, and for the most massive halos $10^{14} < M/M_{\odot} < 5\times 10^{14}$ . Only the case $\alpha=1.5$ is lower than $\sim 10\%$ from the others, which are practically identical.

The chameleon mechanism - the screening mechanism that makes $f(R)$ gravity viable - is not very effective for the parameter choices considered in this paper. This explains the large deviations from $\Lambda$CDM in the observables we considered. Especially the cluster count signatures of $40-100\%$ in the high-mass end $M\gtrsim 10^{14.5}M_{\odot}$ disagree with current observations. This does not rule out the model per se, but require choices of parameters where $|f_{R0}|$ is much smaller today, which implies that the model has no observable signatures in the background evolution (i.e., in the Hubble factor) of the Universe.

\section*{Acknowledgements}
MVS thanks the Brazilian research agency CAPES and the University of Oslo for support and Max B. Grönke and Amir Hammami for useful discussions. HAW is supported by BIPAC and the Oxford Martin School. DFM would like
to thank the Research Council of Norway for funding. The simulations were performed on the NOTUR cluster HEXAGON, which is the computing facility at the University of Bergen.

\bibliographystyle{aa}
\bibliography{bibfile.bib}

\end{document}